# Dendritic Integration Regulation and Neuronal Arithmetic Implemented in a Proton-Coupled Neuron Transistor

Changjin Wan[1,2], Ning Liu[1], Ping Feng[1], Liqiang Zhu[1,2], Yi Shi[1] & Qing Wan[1]

1) School of Electronic Science & Engineering, and Collaborative Innovation Center of Advanced Microstructures, Nanjing University, Nanjing 210093, China

2) Ningbo Institute of Material Technology and Engineering, Chinese Academy of Sciences, Ningbo 315201, China

**Neuron is the most important building block in our brain, and information processing in individual neuron involves the transformation of input synaptic spike trains into an appropriate output spike train. Hardware implementation of neuron by individual ionic/electronic coupled device is of great importance for enhancing our understanding of the brain and solving sensory processing and complex recognition tasks. Here, we provide a proof-of-principle artificial neuron with multiple presynaptic inputs and one modulatory terminal based on a proton-coupled oxide-based electric-double-layer transistor. Regulation of dendritic integration was realized by tuning the voltage applied on the modulatory terminal. Additionally, neuronal gain control (arithmetic) in the scheme of temporal-correlated coding and rate coding are also mimicked. Our results provide a new-concept approach for building brain-inspired neuromorphic systems.**

Email: wanqing@nju.edu.cn

## Introduction

Our brain can perform a simple cognitive task by consuming only ~20 W of power because it processes information using energy-efficient, highly parallel, event-driven architectures as opposed to clocked serial processing[1-3]. At the single neuron level, information processing involves the transformation of input synaptic spike trains into an appropriate output spike train. In the beginning, individual neuron was regarded as the simple linear summation and thresholding device, and even low level computations, such as multiplication of signals, had to be carried out by groups of neurons[4,5]. However, mounting theoretical work suggested that single neuron could act as more powerful computational units[6-8]. Communication between neurons in the brain occurs primarily through synapses made onto elaborate treelike structures called dendrites. Dendritic processing is highly nonlinear, and such dendritic nonlinearities enhance processing capabilities at the single-neuron level [9-14].

The idea of building bio-inspired adaptive solid-state devices has been around for decades. In the past five years, artificial synapses have been demonstrated on a broad spectrum of two terminal devices, such as $Ag_2S$ atom switch, $WO_x$ memristor, $HfO_x$-based resistive switching, and $Ge_2Sb_2Te_5$-based phase change switch [15-22]. Recently, artificial synapses based on three-terminal transistors gated by ionic liquid or solid electrolyte films have also attracted considerable attention. In such synaptic transistors, the voltage pulses applied on the gate are usually regarded as the pre-synaptic spikes and the channel conductance is usually regarded as synaptic weight [21, 21].

Neurons have elaborate dendritic trees that receive thousands of synaptic inputs, and dendritic signal integration is one of the fundamental building blocks of information processing in the brain. In the year of 1992, Shibata et al proposed a new-concept device with many input gates that are capacitively coupled to a floating gate, and named it neuron MOS transistor [23]. The summation of gate voltage signals is carried out by the charge sharing among multiple gate capacitors. Up to now, these Si-based neuron transistors were mainly used for traditional logic circuits and chemical sensors, and neuromorphic behaviors and applications were not investigated. Here, laterally-coupled amorphous oxide-based neuron transistors with multiple presynaptic inputs and a modulatory terminal are proposed based on the electric-double-layer (EDL) modulation. Sublinear to superlinear regulation of spatial summation was realized. What's more, multiplicative algebraic transformation was also experimentally demonstrated in both temporal coding and rate coding schemes. Such oxide-based neuron transistors processed at room temperature provide a new-concept approach for neuromorphic computing systems.

## Results

Figure 1**a** shows the schematic image of a neuron with branched dendrites, which can collect, integrate and modulate the presynaptic inputs and transmit the output spikes to other neurons through the axon. Two types of presynaptic inputs are correlated to such processes: the driving inputs that can make the relevant neuron fire strongly, while the modulating input that can alter the effectiveness of the driving input [24, 25]. Figure 1**b** shows a schematic image of our oxide-based neuron transistor fabricated on a proton conducting phosphorus-doped $SiO_2$ nanogranular electrolyte film. The spiking output currents are measured from the semiconducting indium-zinc-oxide (IZO) channel layer with a fixed voltage ($V_R$). Presynaptic driving input terminals (lateral gate electrodes) are named as $P_1$, $P_2$ ... $P_n$, respectively, and a modulatory terminal is named as $P_m$. The transfer curves ($V_{DS}$=1.5 V) measured with different sweep rates ($\partial V_{GS}/\partial t$) ranged from 1.0 V/s to 0.025 V/s were shown in Fig. 1c. Anti-clockwise hystereses were observed when the gate voltage sweeps from -2.0 V to 2.0 V and then sweeps back. When the sweep rate of $V_{GS}$ is 1.0 V/s, a large hysteresis window of ~1.9 V was observed. When the sweep rate of $V_{GS}$ is reduced to 0.025 V/s, the hysteresis window is reduced to be 0.4 V. At the same time, the maximal output currents ($I_{DS}$ at $V_{GS}$=2.0 V and $V_{DS}$=1.5V) reduce from 339 μA to 22.5 μA, and exhibit an exponential decay relationship with the sweep rate of $V_{GS}$, as shown in Fig. 1d. Such sweep rate dependent output current indicates that the operation mechanism of our transistors is based on the proton-related electric-double-layer (EDL) modulation[26, 27]. Proton migration and relaxation in the nanogranular $SiO_2$ under the applied pulsed gate voltage can modulate the carrier concentration of the IZO channel, which results in the change of the output current ultimately (see Supplementary Note 1). Such process is similar to the generation of postsynaptic current (PSC) in a biological neuron[28, 29]. Excitatory postsynaptic currents (EPSCs) or inhibitory postsynaptic currents (IPSCs) can be triggered by spikes from the presynaptic neurons, which enable the postsynaptic neuron to collectively process PSCs from ~$10^4$ synapses and establish the spatial and temporal correlated functions[3]. The EPSC amplitude, which is the difference between the peak current and the base current, is defined as the response/output of our artificial neuron.

A fundamental computation of neurons is the transformation of incoming synaptic information into specific patterns of synaptic output [30, 31]. An important step of this transformation is dendritic integration, which includes addition of unitary events occurring simultaneously in separate regions of the dendrite arbor (spatial summation) and addition of nonsimultaneous unitary events (temporal summation)[31]. What's more, regulation of such transformation through neural circuit elements could have important implications for the computing and memory-related functions[31-34]. Figure 2**a** schematically shows shunting inhibition, an essential mechanism for regulating the responses of neurons. Excitatory input causes inward postsynaptic current that spreads to the soma, however, when the inhibitory and excitatory inputs are stimulated together, the depolarizing current leaks out before it reaches the soma[35]. In our experiment, dendritic regulation responses were measured by applying a modulatory input (spike or bias) on the modulatory terminal ($P_m$) and a synchronous driving input spikes $V_1$ (1.0 V, 10 ms) on $P_1$. As shown in Fig. 2**b**, when an inhibitory spike $V_m$ (-1.0 V, 10 ms) was applied synchronously, the EPSC was almost eliminated. When the voltage applied on $P_m$ is set to be zero during the measurement, the EPSC is measured to be 113 nA. When an excitatory spike $V_m$ (1.0 V, 10 ms) is applied on the modulatory terminal ($P_m$) synchronously, the EPSC is increased to be 298 nA. Such results indicate that the dendritic output can be suppressed by inhibitory spikes and augmented by excitatory spikes. In other word, neuron transistor output can be regulated by excitatory and/or inhibitory spikes. As shown in Fig. 2**c**, when $P_m$ is biased at different voltages of -0.5, 0 and 0.5 V, different EPSCs of 17 nA, 102 nA and 383 nA are measured, respectively. The response currents against $V_1$ and $V_m$ were also shown as a 2-D surface plot in Fig. 2**d**. The response current increases gradually when $V_1$ and $V_m$ increase. For a medium driving input spike $V_1$ (0.5 V, 10 ms), the curve slopes were 66 nA/V, 134 nA/V and 198 nA/V at $V_m$=−0.5, 0 and 0.5 V, respectively. For a large driving input spike $V_1$ (1.0 V, 10 ms), the curve slopes is tuned to be 277 nA/V, 401 nA/V and 416 nA/V for $V_m$=−0.5, 0 and 0.5 V, respectively. These results indicated that the dynamic range of the neuronal response could be largely tuned by the modulatory voltage.

Figure 3**a** schematically shows the spatial integration regulation by the modulatory input. The red curves and blue curves are the EPSCs ($A_1$, $A_2$) triggered by two spatial isolated presynaptic inputs, respectively. The expected sum ($S_E$, dotted curves) are defined as the arithmetic sum of two individual responses ($A_1+A_2$) and the measured sum ($S_M$, black curves) are the measured responses to the combined stimulus (the combined response). Nonlinearity of spatial summation is called sublinear when the response to two or more inputs is always less than the sum of the individual responses, and called superlinear when the combined response always exceeds the linear prediction [32]. The driving input spikes ($V_1$ and $V_2$) of 1.4 V applied on $P_1$ and $P_2$, respectively triggered individually and then triggered simultaneously, and thus the EPSC amplitudes ($A_1$, $A_2$ and $S_M$, respectively) can be measured successively. The modulatory inputs ($V_m$) applied on $P_m$ triggered synchronously with the driving inputs. All the duration time of the spikes is 10 ms. Figure 3**b** and 3**c** show the regulation results of responses when the modulatory input is -0.5 and 0 V, respectively. When $V_m$= -0.5 V, the $A_1$, $A_2$, and $S_M$ were 37, 19 and 110 nA, respectively, indicating a obviously superlinear summation. When $V_m$=0 V, the $A_1$, $A_2$, and $S_M$ were 182, 132 and 287 nA, respectively,

indicating an approximately linear summation. To evaluate the regulation efficacy on spatial summation, we systematically varied the driving input voltage at the different modulatory inputs of -0.5, 0 and 0.5 V, respectively, and plotted the measured sum versus the expected sum as shown in Fig. 3**d**. The dashed line in the plot denotes exact linear summation. Therefore, when $V_m$= -0.5 V, the output nonlinearity for full range of stimuli was superlinear; when $V_m$= 0 V, the output nonlinearity was nearly linear; when $V_m$= 0.5 V, the output nonlinearity was sublinear. Such results show that the nonlinearity of spatial summation can be regulated by intentionally tuning the modulatory input, indicating the potential implication for neural computing.

Next, the implementation of neuronal arithmetic in individual neuron transistor is discussed. Neuronal arithmetic refers to the multiplicative or additive modulation on input-output relationship by tuning the modulatory input[5, 36-39]. The neuronal coding is often classified into two categories: temporal correlated coding (correlation in the input timing) and rate coding (correlation in the input rate)[40, 41]. Figure 4**a** schematically shows the neuron temporal correlated coding scheme in which a neuron encodes the information through the precise timing of the presynaptic inputs. To reproduce this coding scheme in our oxide-based electronic dendrite, we applied two asynchronous driving inputs on $P_1$ and $P_2$, respectively, and investigated the modulation effects of $P_2$ on $P_1$. When a presynaptic spike (0.5 V, 10 ms) was applied on $P_1$, the EPSC amplitude ($A_0$) was ~210 nA. This EPSC amplitude ($A_1$) can be tuned by a preceded presynaptic spike on $P_2$, as shown in Fig. 4**b**, where the time interval ΔT is 150 ms. When a positive presynaptic spike (0.5 V, 10 ms) on $P_2$ was used, the EPSC amplitude ($A_1$) is ~250 nA (Fig. 4**b**, left), higher than $A_0$. In contrast, when a negative presynaptic spike (−0.5 V, 10 ms) on $P_2$ was used, the EPSC amplitude ($A_1$) is ~180 nA (Fig. 4b, right), lower than $A_0$. To gain more information on the effect of $P_2$ on $P_1$, $A_1/A_0$ was defined as the facilitation/depression ratio and plotted as a function of ΔT, as shown in Fig. 4c. When ΔT was 20 ms, the facilitation ratio was 136%.The facilitation ratio decreased gradually with increasing ΔT. In contrast, the depression ratio was 66%, which increased gradually with increasing ΔT. Both the facilitation and depression ratios were approaching to 100% at very large ΔT, where the relation between the two spikes was very weak. The ratio can be fitted by

$$F(\Delta T) = \frac{A_1}{A_0} = 1 \pm B \cdot \exp[-(\frac{\Delta T - t_w}{\tau})^\beta],$$ where $\tau$, $t_w$, $B$ and $\beta$ are the time constant of proton diffusion in the $SiO_2$ electrolyte, the presynaptic spike width, a proportionality coefficient and a material-dependent parameter(see Supplementary Note 1), and are estimated to be 115 ms, 10 ms, 0.45 and 0.74, respectively. The observed results can be understood by considering the migration of protons in the $SiO_2$ electrolyte and the electrostatic coupling effect of the electric-double-layer. A preceded positive (or negative) presynaptic spike accumulates (or depletes) protons around the dendrite output terminal. After the spike, protons will migrate back to their equilibrium positions gradually, which augment (or weaken) the EPSC amplitude triggered by the following spike, thereby short-term facilitation (or depression) is observed.

Moreover, the short-term facilitation/depression ratio in the neuron transistors can be further tuned by the modulatory input. In the case of temporal correlated coding, the neuronal input-output relationship is defined as the relationship between the dendritic response and the time interval of input spikes. Here, multiplicative output modulation was realized in our electronic dendrite by applying a modulatory input of $V_m$ (−0.1 and 0.1 V) on $P_m$, as shown in

Fig. 4**d**. The slopes of the output curves also increase with $V_m$; for example, the slopes at ΔT=50 ms are −0.37, −0.52 and −0.71 %/ms for $V_m$=−0.1, 0 and 0.1 V, respectively. The output modulation curves can be written as a multiplicative form of $G(V_m) \times F(\Delta T)$, where $G(V_m)=1+k \cdot V_m$ and k is a constant coefficient(see Supplementary Note 2). Systematic changes in the slope or the gain of the neural input-output relationship induced by multiplicative operation underlies a wide range of neural processes, including translation-invariant object recognition, visually guided reaching, collision avoidance, etc. [42-44] The results indicate that our artificial dendrite could be potentially applied in patternrecognition, sensory processing, etc., based on the temporal correlated coding scheme.

Next, we discuss the realization of neuronal arithmetic in rate coding mode in the neuron transistor. Figure 5**a** schematically shows the neuron rate coding scheme. Asynchronous excitatory synaptic input trains with certain mean frequency are triggered from multiple presynaptic neurons. When two Poisson distribution trains (1.0 V, 10 ms) at 20 Hz were applied on $P_1$ and $P_2$, respectively, EPSC amplitude reached the peak value of ~610 nA at the most intensive region of the spike (Fig. 5**b**). With changing the mean frequency of the input trains, the EPSC amplitude can be tuned, as shown in Fig. 5**c**, where the EPSC amplitudes (F(*f*)) were plotted versus the mean frequency (*f*) of the dendritic inputs. The EPSC amplitude increases with increasing the frequency, due to the enhanced coupling at lower time interval between the presynaptic spikes (see Supplementary Note 3). In the case of rate coding, the neural input-output relationship is defined as the relationship between the frequency of the input spike and the dendritic output. As shown in Fig. 5**d**, the EPSC amplitude curves can be tuned by the modulatory input $V_m$ (−0.1 and 0.1 V) on $P_m$. The slopes of the output curves increase with $V_m$; for example, the slopes at $f$= 10 Hz are 20, 22 and 25 nA/Hz for $V_m$= −0.1, 0 and 0.1 V, respectively. The output modulation curves can be expressed as a multiplicative form of $G(V_m) \times F(f)$, where $G(V_m) = 1+k \cdot V_m$ and k is a constant(see Supplementary Note 3). Thus, multiplicative output modulation of rate coding scheme was also realized.

## Discussion

At present, artificial synapses based on two terminal memristors or phase-change memory devices are very pupular for neuromorphic engineering applications [15-20]. Recently, artificial synapses based on three-terminal transistors have also gained considerable attention. It was reported that the advantages of three-terminal transistors over two-terminal resistive switches included reduced sneak path problem and the ability for concurrent signal transmission and long term potential, which is similar to neural circuits and which removes the need for complex circuit timing algorithms[45]. Neuron transistors with multi-gate structure proposed by our group are particularly favorable for artificial neuron constrcution. For complex neuromorphic computation application, high integration density is required. At present, the device size of our proof-of-principle devices is large. In the future, scaling the size of the neuron transistors down to 100 nm should be possible when a advanced photolithography process is used. At the same time, 3D integration of our neuron transistors and fabrication on flexible substrates are also possible because all processes involved in our device fabrication are performed at room temperature. As compared to IZO channel layer, InGaZnO$_4$ channel layer has a very low intrinsic electron concentration, and InGaZnO$_4$-based transistors usually operated in enhancement mode. So InGaZnO$_4$-based EDL transistores can further reduce the power consumption of the artificial neurons.

In conclusion, our investigation presents the first demonstration of oxide-based neuron transistors (artificial neuron) with multi-gate structure based on the EDL modulation. Regulation of dynamic range of dendritic response and spatial summation were experimental demonstrated.What's more, multiplicatively neural output modulation were also realized in such artificial neuron due to the proton-related lateral electric-double-layer modulation. As these functions are highly correlated to neural computations such as pattern recognition, sensory processing, etc., our artificial neurons could be potentially as a building block for neuromorphic computing systems.

**Methods**

**Fabrication of the laterally-coupled neuron transistors**.

Laterally coupled IZO-based transistors were fabricated on glass substrates at room temperature. Firstly, phosphorus-doped nanogranular SiO$_2$ films with a thickness of about 500 nm were deposited on the glass substrate by plasma enhanced chemical-vapor deposition (PECVD) using SiH$_4$/PH$_3$ (95%/5%) mixture and high-purity O$_2$ as the reactive gases. Then, a 30 nm thick patterned indium-zinc-oxide (IZO) channel layer was deposited on the P-doped SiO$_2$ electrolyte films by radio-frequency (RF) magnetron sputtering with the aid of a nickel shadow mask. The sputtering was performed using an IZO ceramic target with a RF power of 100 W and a working pressure of 0.5 Pa. The channel width and length were 240 μm and 80 μm, respectively. Finally, patterned 100-nm thick aluminum (Al) source/drain and gate electrodes (240 μm×200 μm) were deposited through another nickel shadow mask by thermal evaporation method.

**Electrical Measurements**.

Electrical measurements were performed on a semiconductor parameter characterization system (Keithley 4200 SCS) at room temperature and a relative humidity (RH) of 50%. For Excitatory postsynaptic currents (EPSCs) or inhibitory postsynaptic currents measurement, a reading voltage ($V_R$) of 0.5 V is applied between source and drain electrodes to measure the channel current.

**Reference**
1. Markram, H. The blue brain project. *Nat. Rev. Neurosci.* **7**, 153-160 (2006)
2. Machens, C. K. Building the human brain. *Science* **338**, 1156-1157 (2012).
3. Gerstner, W. & Kistler, W. M. Spiking Neuron Models: Single neurons, populations, plasticity. (Cambridge University Press, Cambridge, UK, 2002).
4. Mcculloch, W. S. & Pitts, W. A logical calculus of the ideas immanent in nervous activity. *Bull. Math.Biophys.* **5**, 115–133 (1943).
5. Silver, R. A. Neuronal arithmetic. *Nat. Rev.* **11**, 474-489 (2010).
6. Koch, C., Poggio, T. & Torre, V. Nonlinear interactions in a dendritic tree: localization, timing and role in information processing. *Proc. Natl Acad. Sci.* **80**, 2799–2802 (1983).
7. Williams, S. R. Spatial compartmentalization and functional impact of conductance in pyramidal neurons. *Nat. Neurosci.* **7**, 961-967 (2004).
8. McAdams, C. J. & Maunsell, J. H. R. Effects of attention on orientation-tuning functions of single neurons in macaque cortical area V4. *J. Neurosci.* **19**, 431–441 (1999).
9. Yuste, R. & Denk, W. Dendritic spine as basic functional units of neuronal integration. *Nature* **375**, 682-684 (1995).
10. Branco, T. & Hausser, M. The single dendritic branch as a fundamental functional unit in the nervous system. *Curr.Opin.Neurobiol.* **20**, 494–502 (2010).
11. Polsky, A., Mel, B. M. & Schiller, J. Computational subunits in thin dendrites of pyramidal cells. *Nat. Neurosci.* **7**, 621-627 (2004).
12. Spruston, N. & Kath, W. L. Dendritic arithmetic. *Nat. Neurosci.* **7**, 567-569 (2004).
13. London, M. & Hausser,M. Dendritic computation, A*nnu. Rev. Neurosci.* **28**, 503–532 (2005).
14. Xu., N. et al. Nonlinear dendritic integration of sensory and motor input during an active sensing task. *Nature* **492**, 247-251 (2012).
15. Yu, S. *et al.* A low energy oxide-based electronic synaptic device for neuromorphic visual systems with tolerance to device variation. *Adv.Mater.* **25**, 1774–1779 (2013).

16. Ohno,T. *et al.* Short-term plasticity and long-term potentiation mimicked in single inorganic synapses. *Nat. Mater.* **10**, 591-595 (2011).
17. Pickett, M. D., Medeiros-Ribeiro, G. & Williams R. S. A scalbale neuristor built with Mott memristors. *Nat. Mater.* **12**, 114–117 (2013).
18. Shen, A. M. *et al.* Analog neuromorphic module based on carbon nanotube synapses. *ACS Nano* **7**, 6117-6122 (2013).
19. Chang, T., Jo, S. H. & Lu, W. Short-term memory to long-term memory transition in a nanoscale memristor. *ACS Nano*, **5**, 7669-7676 (2011)
20. Kuzum, D., Jeyasingh, R. G. D., Lee, B. & Wong H.-S. P. Nanoelectronic programmable synapses based on phase change materials for brain-inspired computing. *Nano Lett.* **12**, 2179-2186 (2012).
21. Kim, K., Chen, C. -L., Truong, Q., Shen, A. M. & Chen, Y. A carbon nanotube synapse with dynamic logic and learning. *Adv. Mater.* **25**, 1693-1698 (2013).
22. Zhu, L. Q., Wan, C. J., Guo, L. Q., Shi, Y. & Wan, Q. Artificial synapse network on inorganic proton conductor for neuromorphic systems. *Nat. Commun*. **5**, 3158 (2014).
23. Shibat T. & Ohmi T. A Functional MOS Transistor Featuring Gate-Level Weighted Sum and Threshold Operations. IEEE Transactions on Electron Devices. **39**, 1444 (1992).
24. Sherman, S. M. & Guillery, R. W. On the actions that one nerve cell can have on another: Distinguishing "drivers" from "modulators". *Proc. Natl. Acad. Sci. USA.* **95**, 7121–7126 (1998).
25. Crick F. & Koch C. Constraints on cortical and thalamic projections: the no-strong-loops hypothesis. *Nature* **391**, 245-250, (1998).
26. Panzer, M. J., Newman, C. R., & Frisbie, C. D. Low-voltage operation of a pentacene field-effect transistor with a polymer electrolyte gate dielectric. *Appl. Phys. Lett.* **86**, 103503 (2005).
27. Zhu, L. Q., Sun, J., Wu, G. D., Zhang, H. L. & Wan, Q. Self-assembled dual in-plane gate thin-film transistors gated by nanogranular SiO2 proton conductors for logic applications. *Nanoscale* **5**, 1980–1985 (2013).
28. Bi, G. Q. & Poo, M. M. Synaptic modifications in cultured hippocampal neurons: dependence on spike timing, synaptic strength, and postsynaptic cell type. *J. Neurosci.* **18**, 10464–10472 (1998).
29. Voglis, G. & Tavernarakis, N. The role of synaptic ion channels in synaptic plasticity. *EMBO Rep*. **7**, 1104–1110 (2006).
30. Muller,C., Beck, H., Coulter, D., & Remy, S. Inhibitory Control of Linear and Supralinear Dendritic Excitation in CA1 Pyramidal Neurons. *Neuron* **75**, 851–864 (2012).
31. Magee, J. C. Dendritic integration of excitatory syanptic input. *Nat. Rev. Neurosci*. **1**, 181-190 (2000).
32. Polsky, A., Mel, B. W. & Schiller, J. Computational subunits in thin dendrites of pyramidal cells. *Nat. neruosci.* **7**, 621-627 (2004).
33. Matthew, L. B. *et al.* Regulation of neuronal input transformations by tunable dendritic inhibition. *Nat. Neurosci.* **15**, 423-430 (2012).
34. Isaacson, J. S. & Scanziani, M. How inhibition shapes cortical activity. *Neuron* **72**, 231–243 (2011).
35. Bear, M. F., Connors, B. W. & Paradiso M. A. Neuroscience: Exploring the brain, 3rd

Edition. (Lippincott Williams and Wilkins Press. Philadelphia, USA, 2006).
36. Mitchell, S. J. & Silver, R. A. Shunting inhibition modulates neuronal gain during synaptic excitation. *Neuron* **38**, 433–445 (2003)
37. Salinas, E. & Thier, P. Gain modulation: a major computational principle of the central nervous system. *Neuron* **27**, 15–21 (2000).
38. Ma, W. J., Beck, J. M., Latham, P. E. & Pouget, A. Bayesian inference with probabilistic population codes. *Nat. Neurosci.* **9**, 1432–1438 (2006).
39. Yang, T. & Shadlen, M. N. Probabilistic reasoning by neurons. *Nature* **447**, 1075–1080 (2007).
40. Fuhrmann, G., Segev, I., Markram, H & Tsodyks, M. Coding of Temporal Information by Activity-Dependent Synapses. *J. Neurophysiol.* **87**, 140–148 (2002).
41. Arenz, A., Silver, R. A., Schaefer, A. T. & Margrie, T. W. The contribution of single synapses to sensory representation in vivo. *Science* **321**, 977–980 (2008).
42. Tovee, M. J., Rolls, E. T. & Azzopardi, P. Translation invariance in the responses to faces of single neurons in the temporal visual cortical areas of the alert macaque. *J. Neurophysiol*. **72**, 1049–1060 (1994).
43. Zipser, D. & Andersen, R. A. A back-propagation programmed network that simulates response properties of a subset of posterior parietal neurons. *Nature* **331**, 679–684 (1988).
44. Gabbiani, F., Krapp, H. G., Koch, C. & Laurent, G. Multiplicative computation in a visual neuron sensitive to looming. *Nature* **420**, 320-324 (2002).
45. Ha, S. D., Shi, J., Meroz, Y., Mahadevan, L., & Ramanathan, S. Neuromimetic Circuits with Synaptic Devices Based on Strongly Correlated Electron Systems. *Phys. Rev. Appl.* **2**, 064003 (2014).

**Acknowledgements**

This work was supported by the National Program on Key Basic Research Project (2012CB933004 and 2013CB932900), and the National Natural Science Foundation of China (61425020, 60990314), and a Project Funded by the Priority Academic Program Development of Jiangsu Higher Education Institutions (PAPD), and the Zhejiang Provincial Natural Science Fund (LR13F040001).

**Author contributions**
Q. W. and Y. S. conceived and designed the experiments; C. J. W., P. F., and N. L. fabricated the device and performed device measurement. The manuscript was written by C. J. W., P. F., L. Q. Z., and Q. W.

# Figure caption

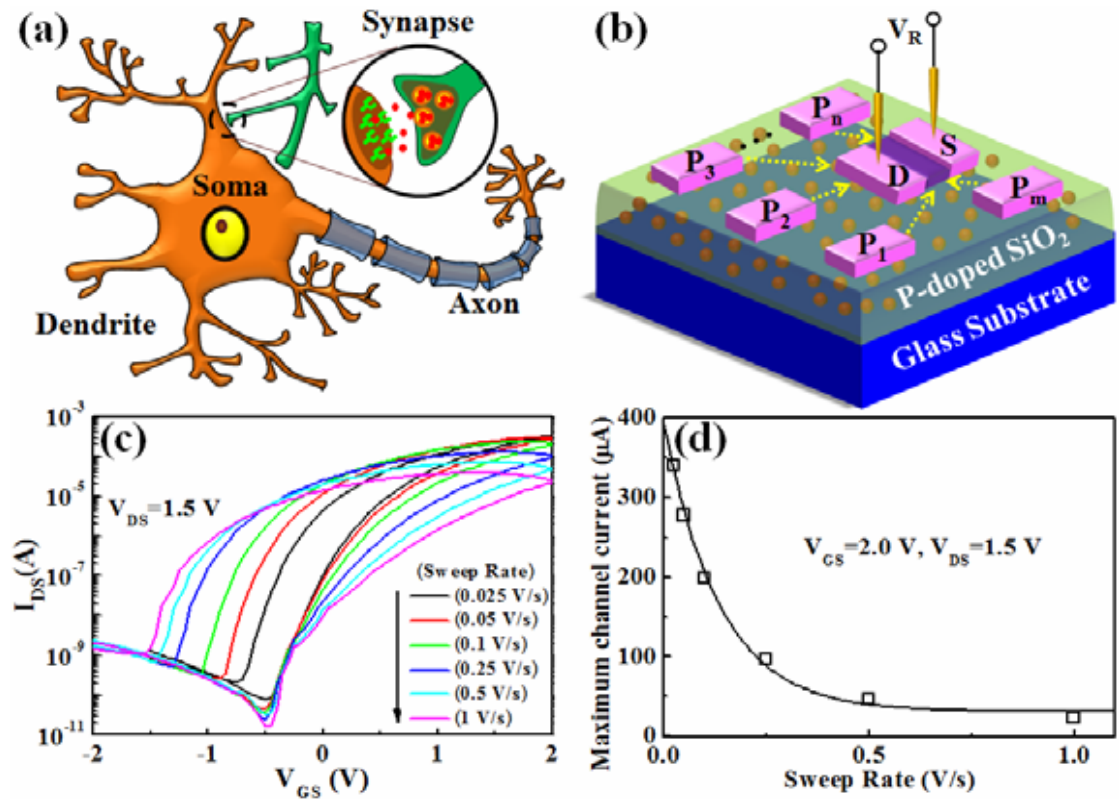

**Figure 1 Artificial neuron with multiple presynaptic inputs and one modulatory input proposed in this work**. **a**, Schematic diagram of a biological neuron with branched dendrites which can collect, integrate and modulate presynaptic inputs through the dendritic spines. The output signal is exported through the axon. **b**, Schematic diagram of an artificial electronic dendrite. The driving inputs are applied on $P_1$, $P_2$ … $P_n$, and the modulatory input is applied on $P_m$. These presynaptic inputs are coupled to the dendritic output (the green pattern) and can trigger EPSCs measured by a reading voltage $V_r$. **c**, Transfer curves at $V_{DS}=1.5$ V with different sweep rate ranged from 0.025 to 1 V/s. **d**, The channel current ($I_{DS}$) at $V_{GS}=2.0$ V was plotted versus the sweep rate ($\partial V_{GS}/\partial t$). The circles were experimental data and the curve was the fitting curve.

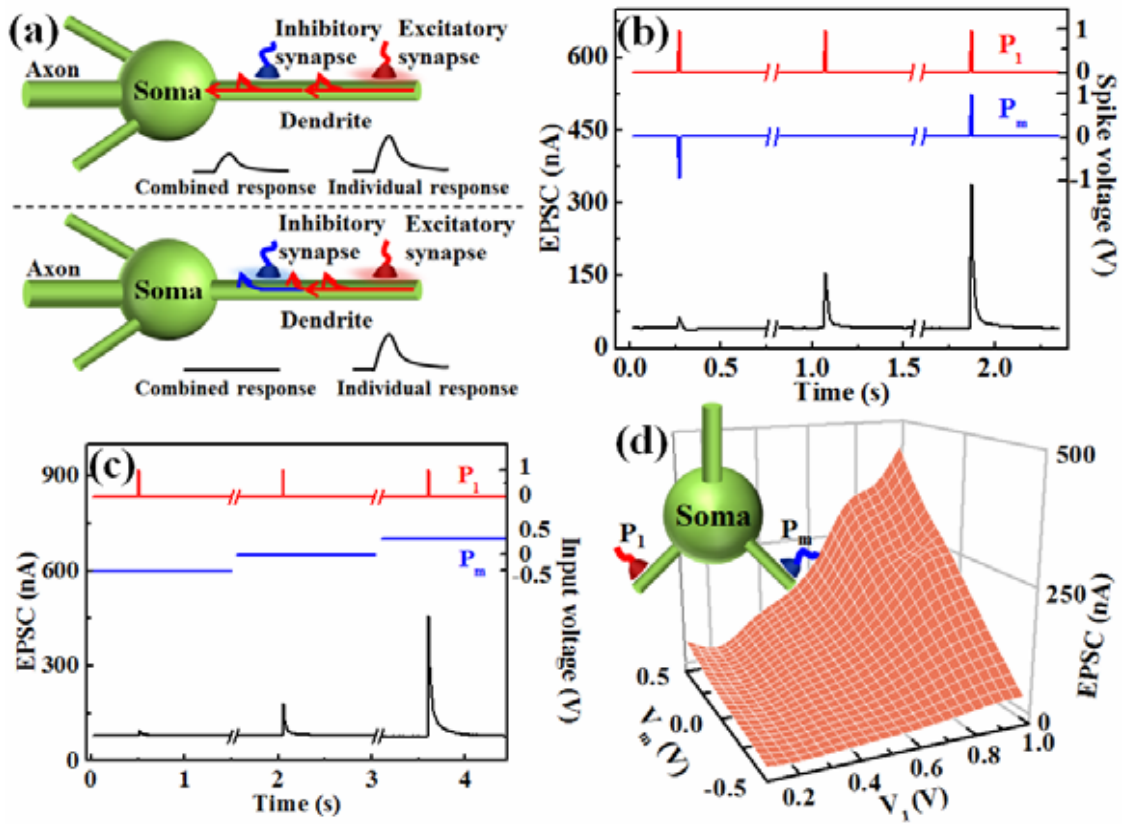

**Figure 2| Regulation of dynamic range of the dendritic response. a,** The schematic diagram of shunting inhibition, an essential mechanism of neural response regulation. **b,** Neural response regulation by applying a modulatory input spike. Modulatory input spike of -1.0, 0 and 1.0 V were triggered in sequence, and three repeated driving inputs (1.0 V, 10 ms) were triggered simultaneously with each modulatory input spike. **c,** Neural response regulation by applying a modulatory input bias. Modulatory input bias of -0.5, 0 and 0.5 V were applied in sequence and three repeated driving inputs (1.0 V, 10 ms) were triggered during each modulatory input bias. **d,** The dynamic range of the dendritic response regulated by modulatory input bias were plotted as a 2-D surface.

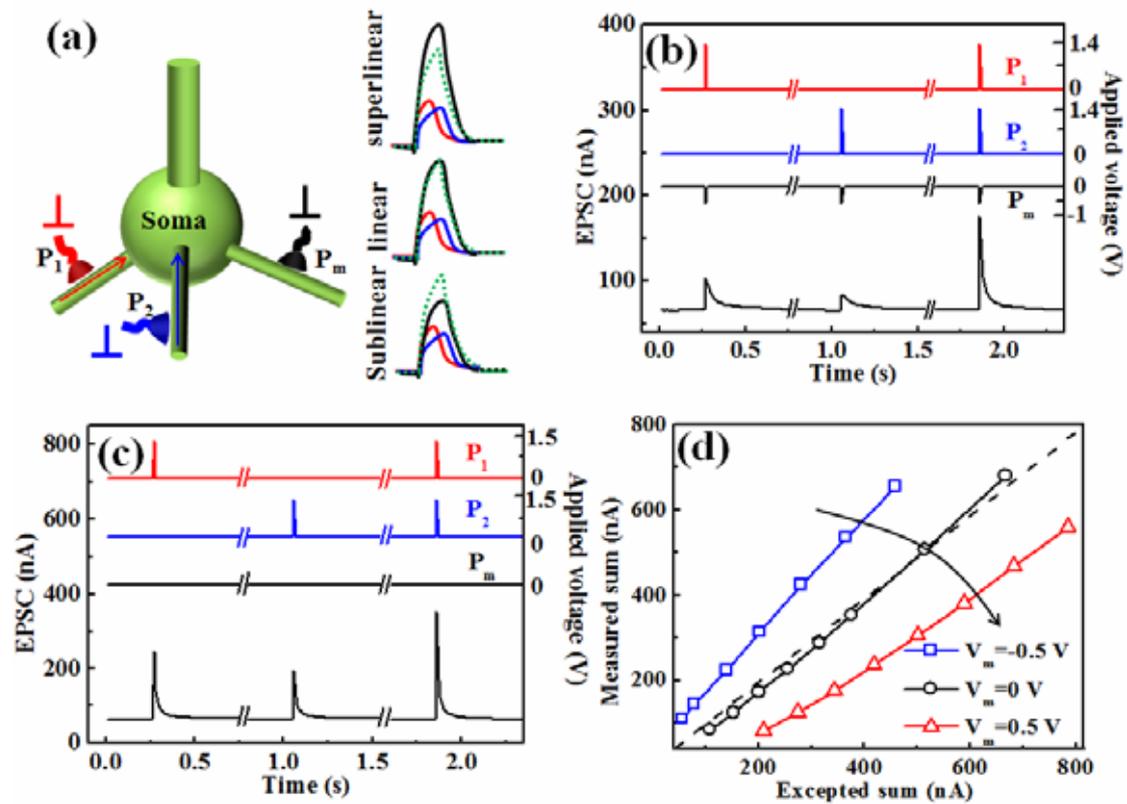

**Figure 3 Regulation of spatial summation nonlinearity. a**, Schematic diagram of regulation of spatial summation. The red line and blue line are EPSCs triggered by two spatial isolated driving inputs ($P_1$, $P_2$). The dotted line is the arithmetic sum of the two EPSCs. The black line is the response when the two inputs triggered simultaneously. **b** and **c**, The driving input spikes (1.4 V, 10 ms) applied on $P_1$ and $P_2$, respectively, were firstly triggered individually and then triggered simultaneously. The modulatory input is triggered simultaneously with the driving inputs, and the magnitudes are -0.5 V (b) and 0 V (c), respectively. **d**, The measured sum versus the expected sum demonstrated the nonlinearity of the spatial summation regulation. When the modulatory inputs were -0.5, 0 and 0.5 V the summation nonlinearity were superlinear, linear and sub linear, respectively.

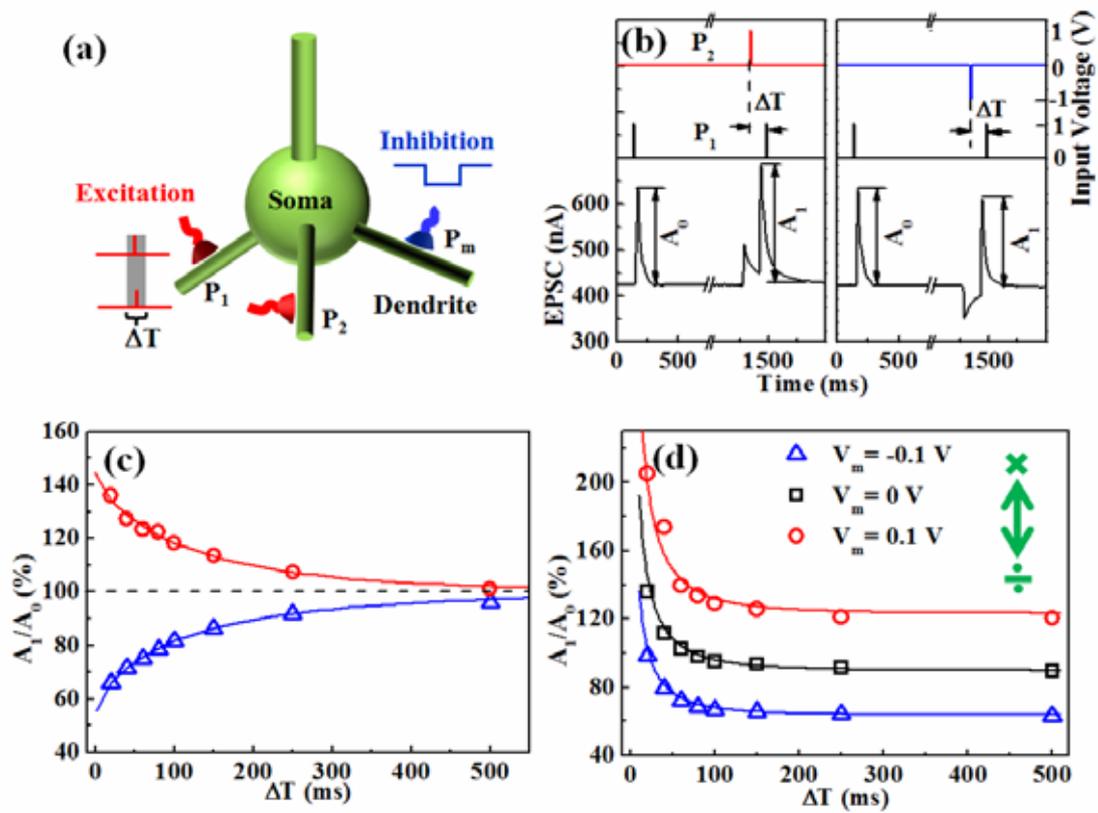

**Figure 4   Neuronal arithmetic in temporal correlated coding scheme demonstrated in our artificial electronic dendrite. a**, Schematic diagram of the temporal-correlated coding neuron model. The red and blue input signals denote the driving and the modulatory inputs, respectively. **b**, EPSCs triggered by a pair of voltage pulses: left panel, two positive pulses; right panel, a negative pulse followed by a positive pulse. The pulses are applied on $P_1$ and $P_2$, respectively. The time interval ΔT is 150 ms. **c**, The Facilitation/depression ratios ($A_1/A_0$) are plotted against the time interval ΔT. The triangles and circles are the experimental data and the lines are the fitting curves. **d**, The facilitation curve plottedin **c** can be multiplecatively tuned by the modulatory input $V_m$ (−0.1 and 0.1 V).

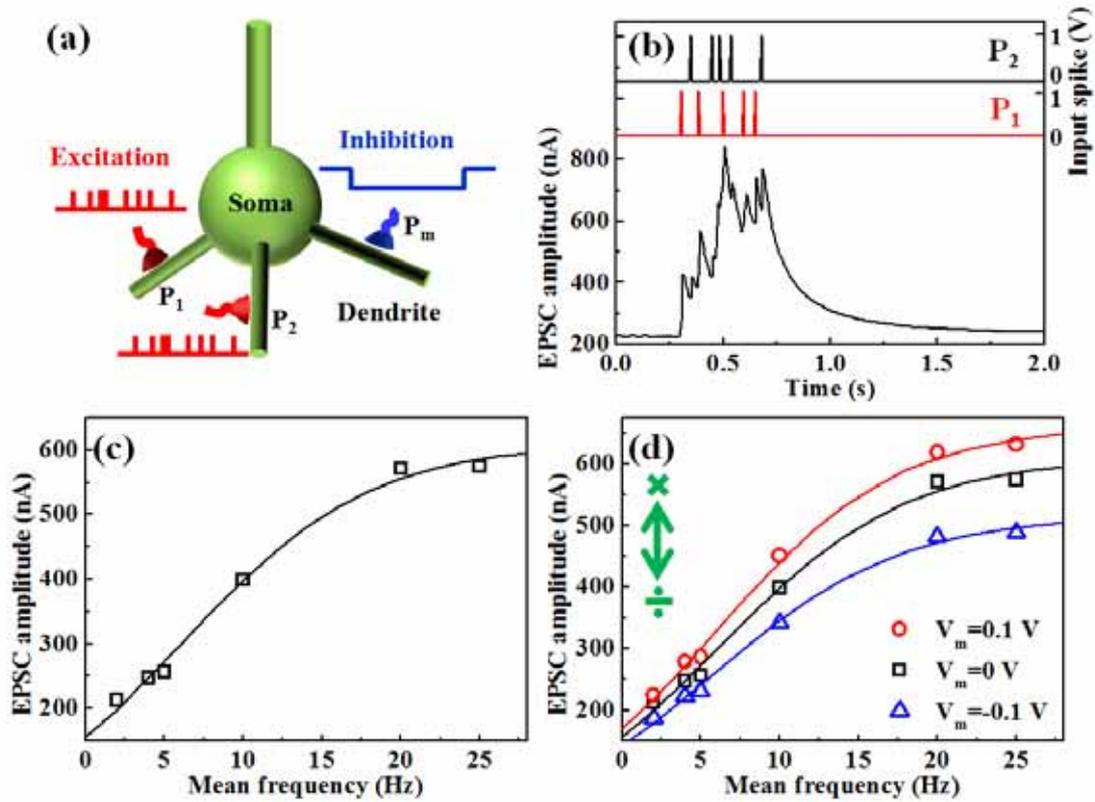

**Figure 5 Neuronal arithmetic in neural rate coding scheme demonstrated in our artificial electronic dendrite. a**, Schematic diagram of the rate coding neuron model. The red and blue input signals denote the driving and modulatory inputs, respectively. **b**, EPSC triggered by two driving input trainson $P_1$ and $P_2$. The input trains arecomposed of spikes (1 V, 10 ms) and obey Poisson distribution with a mean frequency of 20 Hz. **c**, The dendritic outputplotted against the mean frequency of the driving input. The squares are the experimental data and the line is a fitting curve. **d**, The curve in **c** can be multiplicatively tuned by the modulatory input $V_m$ (−0.1 V and 0.1 V).

# Supplementary Information

**Supplementary Note 1: Postsynaptic current simulation**

During the presynaptic spike, the device can be modeled as an ideal resistor-capacitor (RC) circuit.[1] The PSC increases with time and reaches the maximum value (A) at the end of the voltage pulse ($t=t_w$). During the decay process, the current decreases due to a proton relaxation process in the P-dopedSiO$_2$ electrolyte and can be expressed by the stretched-exponential decay model. The stretched-exponential function, also known as Kohlrausch law, can be written as I(t)=I•exp[-(t/$\tau$)$^\beta$], where I, $\tau$ and $\beta$ are the prefactor, the characteristic relaxation time, and the stretch index.[2-4] Therefore, the output current ($I_{DS}$) defined as postsynaptic current (PSC), can be fitted by

$$I_{DS}(t) = A \cdot \exp\left[-\left(\frac{t-t_w}{\tau}\right)^\beta\right] + I_0 \qquad (1)$$

where $A$, $t_w$, $\tau$, $\beta$ and $I_0$ are the PSC amplitude, the voltage pulse width, a time constant, a stretch index and the base current, respectively.

In the PPF case, an EPSC triggered by a presynaptic spike on P$_1$ reaches the maximum of $A$ at the end of the pulse ($t=t_w$). After that, the current loss yields the Eq. (1) and reaches to a value of $A(t=\Delta T) = A \cdot \exp\left[-\left(\frac{\Delta T-t_w}{\tau}\right)^\beta\right]$ when the following pulse with the time interval of ΔT is applied on P$_2$. We assumed such temporal summation is linearity, thus the EPSC amplitude at t=ΔT is $A_1 = A_0 + A(t=\Delta T)$, where $A_0$ is the EPSC amplitude triggered by the presynaptic spike on P$_2$. The amplitude ratio of A$_1$/A$_0$ is

$$F(\Delta T) = \frac{A_1}{A_0} = 1 + B \cdot \exp\left[-\left(\frac{\Delta T-t_w}{\tau}\right)^\beta\right] \qquad (2)$$

Where $B$ is a proportionality coefficient correlated to the weight of the two dendrite inputs.

**Supplementary Note 2: Algebraic transformation of temporal correlated coding**

The data in Fig. **3d** can be well fitted by

$$F(\Delta T) = (1 + k \cdot V_{md}) \cdot \left(1 + B \cdot \exp\left(-\left(\frac{\Delta T - t_w}{\tau}\right)^\beta\right)\right) \quad (3)$$

where k and B are constant coefficients. The right section of the function is deduced from Eq. (2). The fitting parameters of Fig. **3d** are k=0.31, B=10, τ=1.16 ms, β=0.36 and $t_w$=10 ms.

**Supplementary Note 3: Algebraic transformation of rate coding**

The EPSC at $V_m$=0 V can be fitted by a transformative sigmoid function which is applied in many natural processes, such as the learning curves of the complex system. [5] The function is

$$F(f) = \frac{A}{(\alpha + \beta \cdot \exp(-\gamma \cdot (f - \delta)))} \quad (3)$$

where κ, α, β, γ and δ are constant coefficients, respectively, and $f$ is the mean frequency of the Poisson-distributed train. The data in Fig. 4c can be well fitted by Eq. (3) with A=2.55×10$^{-7}$, α=0.33, β=0.81, γ=0.17 and δ=0.81.

By applying modulatory input, as the case in Fig. 4d, the data can be fitted by

$$F(f) = \frac{A \cdot (1 + k \cdot V_m)}{(\alpha + \beta \cdot \exp(-\gamma \cdot (f_{dr} - \delta)))} \quad (4)$$

with A=2.55×10$^{-7}$, k=0.77, α=0.33, β=0.81, γ=0.17 and δ=0.81.

## Supplementary References

[1] Yuan, H. T., Shimotani, H., Ye, J. T., Yoon, S., Aliah, H., Tsukazaki, A., Kawasaki, M.& Iwasa Y. Electrostatic and Electrochemical Nature of Liquid-Gated Electric-Double-Layer Transistors Based on Oxide Semiconductors. *J. Am. Chem Soc.* **132**, 18402–18407 (2010).

[2] Scher, H., Shlesinger, M. F.,&Bendler, J. T. Time-scale invariance in transport and relaxation. *Physics Today.* **44**, 26-34 (1991).

[3] Kakalios, J., Street, R. A.,& Jackson, W. B. Stretched-exponential relaxation arising from dispersive diffusion of hydrogen in amorphous silicon. *Phys. Rev. Lett.* **59**, 1037 (1987).

[4] Chang, T., Jo, S. H. & Lu, W. Short-term memory to long-term memory transition in a nanoscale memristor. *ACS Nano* **9**, 7669-7676 (2011).

[5] Little, W. A., Shaw, G. L. Analytic Study of the Memory Capacity of a Neural Network. Math. Biosci. **39**, 281, (1978).